\documentclass[twocolumn,showpacs,superscriptaddress,prb,floatfix]{revtex4}
\usepackage{graphicx,amsmath,amsfonts,bm}
\usepackage[colorlinks=true, citecolor=blue, linkcolor=blue, urlcolor=blue]{hyperref}
\usepackage[all]{hypcap}

\graphicspath{{figures/}}
\def\sgn{\mathop{\rm sgn}\nolimits} 

\begin{document}
\author{Christophe De Beule}
\email{christophe.debeule@ua.ac.be} \affiliation{Department of Physics, University of Antwerp, 2020 Antwerp, Belgium}
\author{Bart Partoens}
\email{bart.partoens@ua.ac.be} \affiliation{Department of Physics, University of Antwerp, 2020 Antwerp, Belgium}
\date{\today}
\title{Gapless interface states at the junction between two topological insulators}

\begin{abstract}
We consider a junction between two topological insulators, and calculate the properties of the interface states with an effective low energy Hamiltonian for topological insulators with a single cone on the surface. This system bears a close resemblance to bilayer graphene, as both result from the hybridization of Dirac cones. We find gapless interface states not only when the helicity direction of the topological surface states are oppositely oriented, but they can also exist if they are equally oriented. Furthermore, we find that the existence of the interface states can be understood from the closing of the bulk gap when the helicity changes orientation. Recently, superluminal tachyonic excitations were also claimed to exist at the interface between topological insulators. However, here we show that these interface states do not exist.
\end{abstract}

\maketitle

\section{Introduction}

Topological insulators (TIs) \cite{kane1,bern,konig,fu1,moore1,fu2,mura1,hsieh,zhang,roy,xia,chen} are a newly discovered class of materials, that have attracted a lot of interest from the condensed matter community in the last few years \cite{kane2,qi1,moore2,hasan,qi2,mura2}. The strong three-dimensional TI, e.g. Bi$_2$Se$_3$, is insulating in the bulk with gapless surface states protected by time-reversal (TR) symmetry as a consequence of band inversion by strong spin-orbit coupling (SOC). Elastic backscattering of these states is forbidden due to TR symmetry, and they remain gapless for any TR invariant perturbation of the system that does not close the gap. At low energies, the surface states of Bi$_2$Se$_3$ are given by a single Dirac cone, i.e. they have linear dispersion and a helical spin texture \cite{zhang,xia}.

Recently it was shown that there should also exist protected gapless interface states, based on symmetry arguments, at the junction between two TIs with opposite helicity direction \cite{taka}. These interface states are not protected in the same manner as the surface states of a single TI, instead they are protected by mirror symmetry. The physics of this system resembles certain aspects of bilayer graphene, because both result from the hybridization of Dirac cones.

In this paper, we make a systematic study of the possible combinations of TIs using a quantitative model for a strong TI \cite{zhang,liu1}, and we show the existence of different types of interface states. This model can be derived from $\bm{k}\cdot\bm{p}$ perturbation theory in which the full Hamiltonian is projected on the subspace of states that dominate near the $\Gamma$ point. 
We follow the same approach as a very recent paper \cite{apal} that claims the existence of tachyonlike interface states in this system. We found, however, that this model does not predict tachyonlike solutions. The reason is technical, and due to a wrong implementation of the model under certain pathological conditions.

The paper is further organized as follows. In Sec.\ \ref{sec:model} we briefly discuss the model that is used to describe the TIs, and we explain how the interface states were calculated. Next, we present and discuss our results in Sec.\ \ref{sec:results}, and we prove that the tachyonlike interface states are not physical solutions.
\begin{figure}
  \centering
  \includegraphics[width=0.45\textwidth]{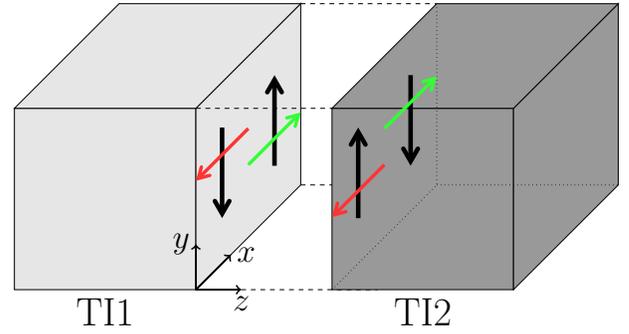}
  \caption{(color online). Junction between TI1 and TI2 with opposite helicity direction. The spin of the surface Dirac cone lies in the plane, perpendicular to the direction of propagation (given here by the $y$-axis), oriented along the positive (green) and negative (red) $x$-axis for TI1 and TI2 respectively.} \label{fig:junction} 
\end{figure}

\section{Model} \label{sec:model}

\subsection{General}

The low energy physics of a bulk strong TI, like Bi$_2$Se$_3$, is described up to order $k^2$ by the effective Hamiltonian \cite{zhang,liu1}
\begin{equation} \label{eq:ham}
  \mathcal{H}(\bm{k},k_z)=\varepsilon+
  \begin{pmatrix}
    M\sigma_z + Bk_z\sigma_x & Ak_-\sigma_x \\
    Ak_+\sigma_x & M\sigma_z - Bk_z\sigma_x
  \end{pmatrix},
\end{equation}
where $\bm{\sigma}$ are the Pauli matrices for the Bi and Se sites, and
\begin{align*}
  \varepsilon(k,k_z)&=C_0+C_1k_z^2+C_2k^2, \\
  M(k,k_z)&=M_0+M_1k_z^2+M_2k^2,
\end{align*}
with $\bm{k}=(k_x,k_y)$ and $k_\pm=k_x\pm ik_y$. The Schr\"odinger equation is given by $\mathcal{H}\Phi=E\Phi$, where $\Phi=\left(\mbox{Bi}_\uparrow,\mbox{Se}_\uparrow,\mbox{Bi}_\downarrow,\mbox{Se}_\downarrow\right)$ gives the amplitude of the spin-orbit coupled $p_z$ orbitals that mainly contribute to the low energy physics of Bi$_2$Se$_3$. The band inversion which characterizes the topological phase, is determined by the condition $M_0M_1<0$ (Ref.\ \onlinecite{liu1}). 

Surface states obtained from this model have linear dispersion near the $\Gamma$ point with Fermi velocity $v=(|A|/\hbar)\left[1-(C_1/M_1)^2\right]^{1/2}$, and a helical spin texture \cite{shan}. The helicity direction of the spin texture depends on the relative sign of the parameters $A$ and $B$, i.e.\ $\left<s_{x,y}\right>\sim\sgn\left({AB}\right)k_{y,x}$ (Ref.\ \onlinecite{liu1}).

\subsection{Junction}

We consider a junction at $z=0$ between two TIs, TI1 ($z<0$) and TI2 ($z>0$) (Fig.\ \ref{fig:junction}). The junction breaks translation symmetry in the $z$-direction, and we let $k_z\rightarrow-i\partial_z$ to obtain a system of second order homogeneous differential equations $\mathcal{H}(\bm{k},-i\partial_z)\Phi=E\Phi$ with $\Phi=e^{i\bm{k}\cdot\bm{\rho}}\bm{\phi}(z)$. This is solved with the ansatz $\bm{\phi}(z)\sim e^{\lambda z}\Psi(E,\bm{k})$, where $\Psi(E,\bm{k})$ is an eigenvector of the Hamiltonian. We obtain a system of algebraic equations that has a nonzero solution if $|\mathcal{H}(\bm{k},-i\lambda)-E|=0$. Details on the solution method are given in the Appendix. Solving for $\lambda$ yields four doubly degenerate $\lambda_\alpha(E,\bm{k})$ $(\alpha=1,\ldots,4)$ in general, each corresponding to two eigenvectors $\Psi_{s,\alpha}$ $(s=1,2)$ (Eq.\ (\ref{eq:vec})). The total solution is then given by the linear combination
\begin{equation} \label{eq:wf}
  \bm{\phi}(z)=\displaystyle\sum\limits_s\displaystyle\sum\limits_\alpha C_{s,\alpha}e^{\lambda_\alpha z}\Psi_{s,\alpha},
\end{equation}
where the coefficients $C_{s,\alpha}(E,\bm{k})$ are found from the boundary conditions.

We want to study states localized at the junction, and therefore we only consider the two $\lambda^{(m)}$ with Re~$\lambda^{(1)}>0$ for TI1 and Re~$\lambda^{(2)}<0$ for TI2, to construct the wave function (\ref{eq:wf}) (see Eq.\ (\ref{eq:lambda})). Here $m$ labels TI1 and TI2 respectively. The remaining boundary conditions are given by the continuity of the wave function and the current at the junction \cite{ben}
\begin{equation} \label{eq:boundary}
  \begin{aligned} 
    \phi^{(1)}(0)&=\phi^{(2)}(0), \\
    \left.\left.\frac{\delta\mathcal{H}^{(1)}}{\delta k_z}\right|_{k_z=-i\partial_z}\phi^{(1)}\right|_{z=0}&=\left.\left.\frac{\delta\mathcal{H}^{(2)}}{\delta k_z}\right|_{k_z=-i\partial_z}\phi^{(2)}\right|_{z=0}.
  \end{aligned}
\end{equation}
The nonzero solutions $C_{s,\alpha}^{(m)}(E,\bm{k})\neq0$ of this system of homogeneous equations, define the dispersion and wave functions of the interface states. We numerically solved this system on a $(E,\bm{k})$ grid by means of the condition number and singular value decomposition of the coefficient matrix.

\section{Results} \label{sec:results}

The parameters of the model (\ref{eq:ham}) are taken from Ref.\ \onlinecite{zhang} and were obtained from \emph{ab initio} calculations of Bi$_2$Se$_3$. We want to study interface states between TIs with opposite helicity, and therefore we focus only on the parameters $A^{(2)}$ and $B^{(2)}$, which determine the helicity direction of the surface states of TI2. We find gapless interface states if the helicity direction of the surface states of the TIs is equally as well as oppositely oriented. Specifically, we find solutions if the sign of the parameters $A$ or $B$ is opposite for the TIs (case 1 and 2), corresponding to TIs with opposite helicity, and also if the sign of both parameters is opposite (case 3), corresponding to TIs with equal helicity. We find no solutions for the other case in which both TIs have equal helicity, i.e.\ if both parameters have the same sign. To simplify the calculation and because the Hamiltonian (\ref{eq:ham}) has full rotation symmetry about the $z$-axis, we take the wavevector along the $x$-direction.
\begin{figure}
  \centering
  \includegraphics[width=0.45\textwidth]{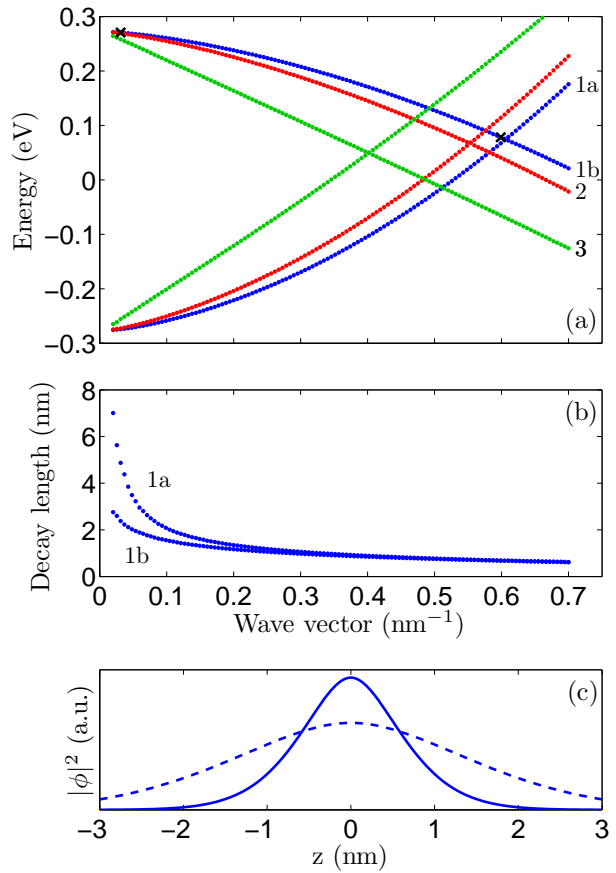}
  \caption{(color online). (a) Dispersion of the interface states for $B^{(2)}=B^{(1)}$, and $A^{(2)}=-A^{(1)}$ (1), $A^{(2)}=-2A^{(1)}$ (2) and $A^{(2)}=-10A^{(1)}$ (3). (b) Decay length of the interface states for the two bands from (a) with $A^{(2)}=-A^{(1)}$. (c) Density of the states of band 1b from (a) marked with a cross, near the crossing point (solid), and closer to the bulk region (dashed).} \label{fig:plot1}
\end{figure}

\subsection{Case 1}

First we consider the case where $B^{(2)}=B^{(1)}$ and $A^{(1)}A^{(2)}<0$. This corresponds with TIs that have surface states with opposite helicity direction (Fig.\ \ref{fig:junction}), and the magnitude of $A^{(1,2)}$ determines the Fermi velocity of the surface states. In Fig.\ \ref{fig:plot1}(a) the dispersion of the interface states is shown for three different values of $A^{(2)}$. The interface states have linear dispersion (in the radial direction) around $k\neq0$ which results from the hybridization of the Dirac cones from the separated TIs, similar to AA stacked bilayer graphene \cite{lob}. The cones intersect closer to the center if their Fermi velocity is smaller. From the energy scale in Fig.\ \ref{fig:plot1}(a) one can see that the conduction/valence band of the cone is pushed into the bulk region of the band structure, which has a band gap around $0.3$ eV \cite{zhang, xia}. Correspondingly, as the bands approach $k=0$, the decay length $\equiv\max\left(1/|\mbox{Re}~\lambda_\alpha|\right)$ diverges exponentially, as shown in Fig.\ \ref{fig:plot1}(b). 

The total density near the junction is shown in Fig.\ \ref{fig:plot1}(c) for a state close to the crossing point and a state close to the bulk region. Both states are marked with a cross in Fig.\ \ref{fig:plot1}(a). We see that the character of the individual surface states is completely lost and the interface states are spread over the entire junction.
\begin{figure}
  \centering
  \includegraphics[width=0.45\textwidth]{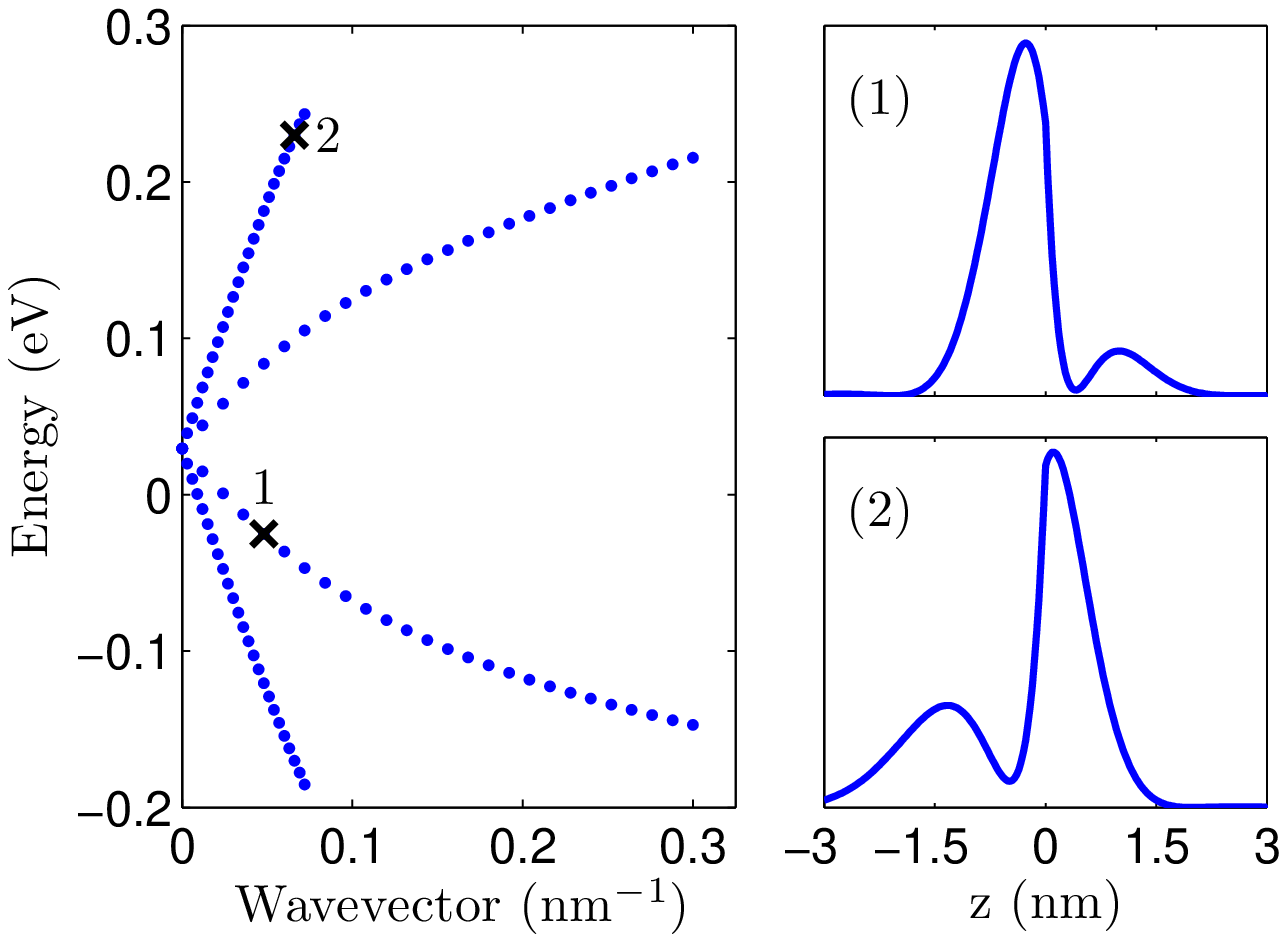}
  \caption{Dispersion of the interface states between TIs for $B^{(2)}=-B^{(1)}$ and $A^{(2)}=10A^{(1)}$. The density (arb.\ units) is shown on the right for the two states that are marked with a cross on the dispersion.} \label{fig:plot2}
\end{figure}

\subsection{Case 2}

Next we consider the case where $B^{(1)}B^{(2)}<0$ and $A^{(1)}A^{(2)}>0$. Again this corresponds with TIs that have surface states with opposite helicity direction (Fig.\ref{fig:junction}), and the magnitude of $A^{(1,2)}$ determines the Fermi velocity of the surface states. In Fig.\ \ref{fig:plot2} the dispersion of the interface states is shown for $B^{(2)}=-B^{(1)}$ and $A^{(2)}=10A^{(1)}$. Unlike case 1, there is little interaction between the surfaces, and the cones lie on top of each other.

The density is also shown for two states in Fig.\ \ref{fig:plot2}, one on each cone. We see that the states on each cone are not localized at the junction, but instead they are localized inside one of the TIs, as we would expect if there is little interaction, and the individual surface states keep most of their original character. TI2 has the largest Fermi velocity, correspondingly the states on the steepest cone are localized in TI2. The Fermi velocity of the cones in the interface spectrum is two times as large as the original cone of the separated TI for TI1 and about 0.8 times smaller for TI2, and thus the cones are attracted to each other.

\subsection{Case 3}

\begin{figure}
  \centering
  \includegraphics[width=0.45\textwidth]{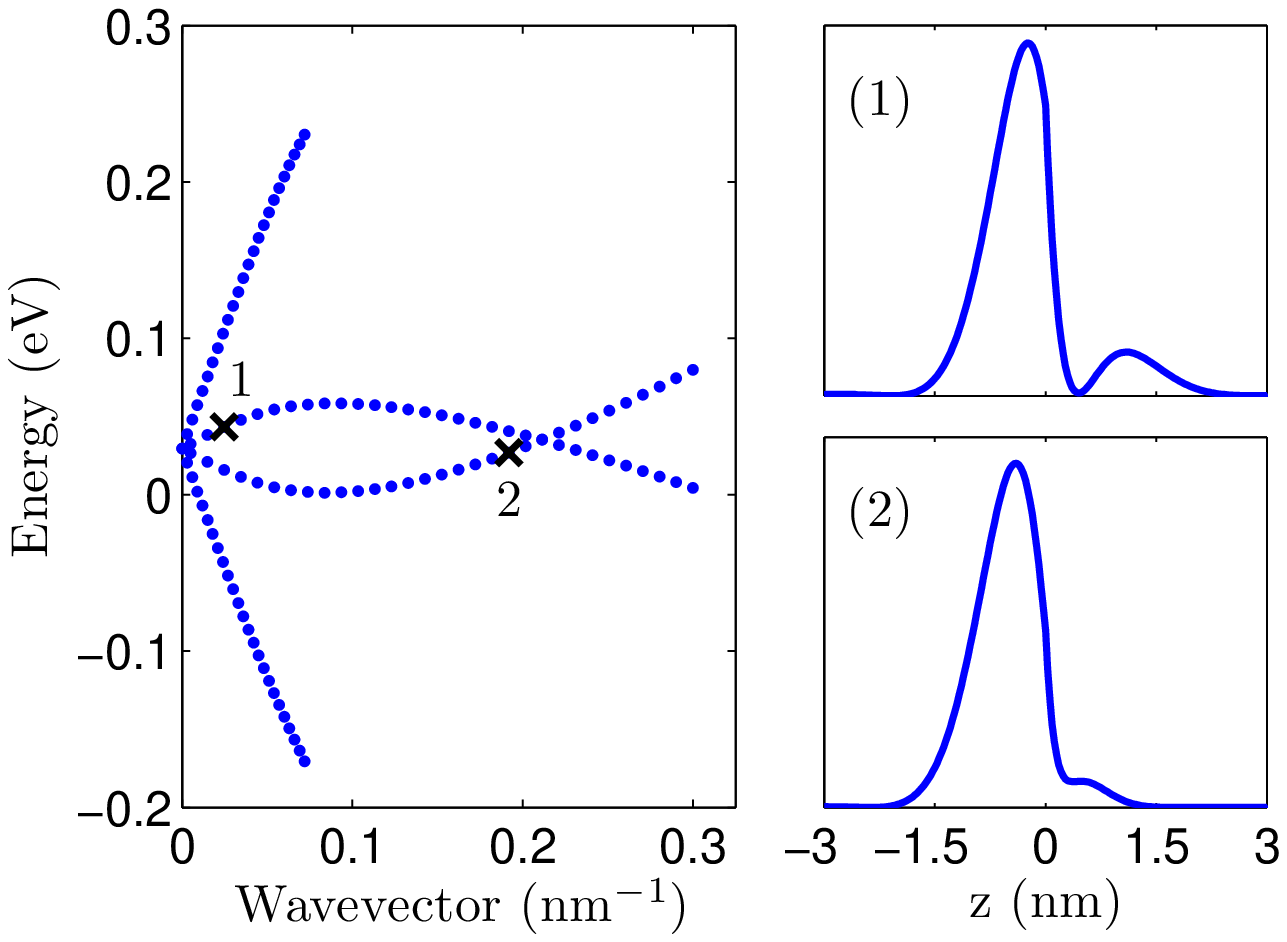}
  \caption{Dispersion of the interface states between TIs for $B^{(2)}=-B^{(1)}$ and $A^{(2)}=-10A^{(1)}$. The density (arb.\ units) is shown on the right for the two states that are marked with a cross on the dispersion.} \label{fig:plot3}
\end{figure}

Finally we consider the case where $B^{(1)}B^{(2)}<0$ and $A^{(1)}A^{(2)}<0$. Unlike the previous two cases, this corresponds with TIs that have surface states with equal helicity direction. The dispersion of the interface states is shown in Fig.\ \ref{fig:plot3} for $B^{(2)}=-B^{(1)}$ and $A^{(2)}=-10A^{(1)}$. We see that we get a combination of the previous two cases, a linear spectrum near the center and a crossing away from the center. The density is also shown for two states in Fig.\ \ref{fig:plot3}, localized on TI1, one near the center and one near the crossing point. We see that near the center, we have the same behavior as in case 2, and closer to the crossing point the interaction between the surfaces is stronger as indicated by the shift of the density towards the junction. The states on the steep cone do not differ much from those of the steep cone of Fig.\ \ref{fig:plot2} and are localized within TI2.

\subsection{Discussion} \label{sec:discussion}

\begin{figure}
  \centering
  \includegraphics[width=0.25\textwidth]{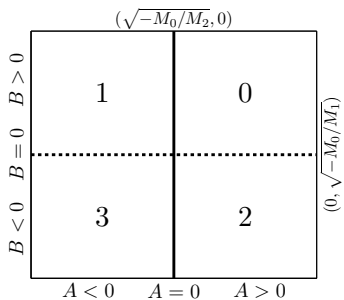}
  \caption{Schematic phase diagram as function of $A$ and $B$. When either of the parameters changes sign, the gap closes at $(k,k_z)$ shown and the helicity changes orientation. Crossing from region 0 to region 1-3 corresponds with case 1-3.} \label{fig:diag}
\end{figure}
If the helicity direction of the TIs is opposite, the existence of the gapless interface states can be understood from scattering of surface states at the junction between the two TIs, as shown in Refs.\ \onlinecite{taka} and \onlinecite{dip}. At normal incidence, the incoming state on the surface of TI1 cannot scatter backwards because of spin conservation. This is similar to graphene, where pseudospin conservation results in Klein tunneling through potential barriers at normal incidence \cite{kats}. However, when TI2 has opposite helicity, the incoming state cannot go forward because there are no states in TI2 that conserve both spin and momentum. This paradox is solved by the existence of gapless interface states between the TIs. At normal incidence the incoming state has to go into the interface. Consequently, these states should exist for $k_x=0$ or $k_y=0$. 

The origin of the surface states in TIs can be understood from the closure of the bulk band gap if the Hamiltonian is transformed continuously from a trivial insulator to a TI. In the same way, we can understand the interface states at the junction between two TIs. We observe that the gap of the bulk band structure of the Hamiltonian (\ref{eq:ham}) closes at $k\neq0$ if the parameter $A$ changes sign, and it closes at $k=0$ if the parameter $B$ changes sign, which could be an indication of a topological phase transition. The origin of the interface states within the model can then also be understood from this closing of the bulk gap when $A=0$ at $k\neq0$ for case 1, and when $B=0$ at $k=0$ for case 2. When both parameters $A$ and $B$ change sign as in case 3, we end up with two TIs with equal helicity direction. Therefore, at first sight it might be surprising that interface states appear. However, from the previous discussion it is clear that the bulk gap closes at two different $k$ values when both $A$ and $B$ change sign, which explains why the resulting spectrum in case 3 (Fig.\ \ref{fig:plot3}) is a combination of the spectra of case 1 (Fig.\ \ref{fig:plot1}(a)) and case 2 (Fig.\ \ref{fig:plot2}). This is illustrated with a schematic phase diagram in Fig.\ \ref{fig:diag}, where we show the four different sign combinations of $A$ and $B$, and the $(k,k_z)$ points at which the gap closes when the helicity changes orientation.

Because of the invariance of the Hamiltonian (\ref{eq:ham}) under any rotation about the $z$-axis, we find gapless interface states along any $\bm{k}$-direction. If we take into account terms up to order $k^3$, the full rotation symmetry is relaxed to threefold $C_3$ rotation symmetry of the Bi$_2$Se$_3$ crystal \cite{liu1}, and we expect a gap to open in the interface spectrum along $k_x\neq0$ or $k_y\neq0$ directions, giving a total of six Dirac cones over the entire Brillouin zone. However, the solution method becomes unpractical in this case, because the equation for $\lambda$ (\ref{eq:lambda}) contains odd powers, and we need to find the roots of a depressed quartic equation. 

Now raises the question if materials can exist for which the parameters $A$ and $B$ changes sign with respect to their Bi$_2$Se$_3$ values. In Ref.\ \onlinecite{liu1} it is argued that the sign of $A$ is determined by the sign of the atomic SOC, while the sign of $B$ is independent of the atomic SOC. However, the atomic SOC parameter, given by the diagonal matrix element of $\hat{H}_{SO}=(2m_e^2c^2r)^{-1}(\partial U/\partial r)\bm{L}\cdot\bm{S}$, does not change sign for Bi$_2$Se$_3$-like TIs because the potential is always attractive for atoms \cite{liu1}. Therefore, the helicity direction is always the same within the Bi$_2$Se$_3$ family of TIs. However, the model (\ref{eq:ham}) is valid for a general TI, and therefore we believe that the interface between TIs with opposite helicity is possible in principle, as was also mentioned in Ref.\ \onlinecite{taka}. Furthermore, there are other mechanisms besides SOC which can give rise to a band inversion. For example, strained bulk HgTe is a 3D TI \cite{brune} in which the band inversion is caused by other relativistic corrections \cite{berch}. Also in type-II InAs/GaSb heterostructures, one can achieve a band inversion from confinement \cite{liu2}.

\subsection{Tachyonlike interface states}

\begin{figure}
  \centering
  \includegraphics[width=0.45\textwidth]{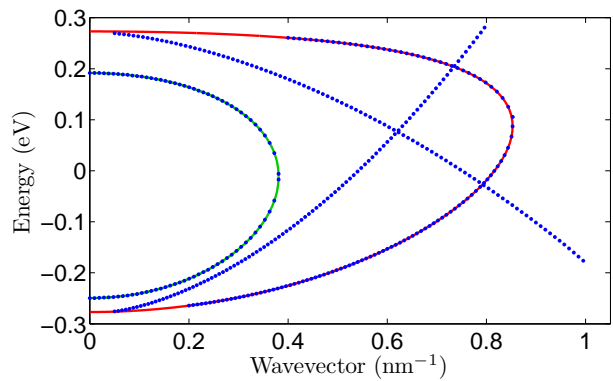}
  \caption{(color online). Dispersion of the interface states for $A^{(2)}=-A^{(1)}$ and $B^{(2)}=3.0$ eV\AA, where we have included the tachyonlike solutions; this corresponds to Fig.\ 3(a) of Ref.\ \onlinecite{apal}. Eq.\ (\ref{eq:tachyon}) is plotted for TI1 (red) and TI2 (green) together with the numerical results (dots). From Eq.\ (\ref{eq:tachyonk}) we find $k_t^{(1)}\approx0.0853$ and $k_t^{(2)}\approx0.0381$.} \label{fig:tachyon}
\end{figure}
We found no signature of tachyons in this system as claimed in Ref.\ \onlinecite{apal}. The reason is the following. If the $\lambda$'s used to construct the general solution (\ref{eq:wf}) become four times degenerate, the solution is no longer valid because it contains the same contribution twice. In this case, the matrix of the homogeneous system, defined by the boundary conditions (\ref{eq:boundary}), is always singular and one seemingly finds solutions. The extra degeneracy occurs at $(E,\bm{k})$ points where the equation in $\lambda^2$ has a repeated root (Eq.\ (\ref{eq:lambda})). This happens when the discriminant vanishes which allowed us to obtain an expression for the `dispersion',
\begin{equation} \label{eq:tachyon}
  \begin{aligned}
    E_t(k)=&\left[C_0-\frac{C_1}{M_1}\left(M_0+\frac{B^2}{2M_1}\right)\right]+\left(C_2-\frac{M_2}{M_1}C_1\right)k^2 \\
    &\pm\frac{1}{2M_1^2}\sqrt{(C_1^2-M_1^2)f(k)},
  \end{aligned}
\end{equation}
with
\begin{equation*}
  f(k)=B^4+4B^2M_0M_1+4M_1(B^2M_2-A^2M_1)k^2.
\end{equation*}
When the argument of the square root becomes negative, the `group velocity' diverges and $E_t$ becomes complex. This happens at $f(k_t)=0$, which gives the tachyonic point \cite{apal}
\begin{equation} \label{eq:tachyonk}
  k_t=\frac{|B|}{2}\sqrt{\frac{B^2+4M_0M_1}{M_1(A^2M_1-B^2M_2)}}.
\end{equation}
In Fig.\ \ref{fig:tachyon} we show the dispersion for the case which was also considered in Ref.\ \onlinecite{apal}, $A^{(2)}=-A^{(1)}$ and $B^{(2)}=3.0$ eV\AA. We see that Eq.\ (\ref{eq:tachyon}) exactly fits the tachyonlike `dispersions' obtained with a numerical calculation that does not take the degeneracy into account. In this case we get two tachyonlike `dispersions' because the degeneracy occurs for $\lambda^{(1)}$ and $\lambda^{(2)}$ for this parameter set. We know that the degeneracy occurs on a real $(E,k)$ grid if $k_t$ is real. The same problem occurs if we consider a vacuum interface, because the degeneracy of $\lambda$ is independent of the boundary conditions. In this case, one can analytically show that the tachyonic solutions are not physical.

The correct general solution at the degeneracy is given by Eq.\ (\ref{eq:wf2}) in the Appendix. Now we only find solutions at the points of intersection between the unphysical tachyonlike `dispersion' and physical solutions.

\section{Conclusion}

In conclusion, we have studied interface states at the junction between two TIs with equally and oppositely oriented helicity direction. The origin of the gapless interface states can be understood from the closing of the bulk gap in the transition from one orientation of helicity to the opposite orientation at different wavevectors. Even if we relax the full rotation symmetry of the model, there should still be gapless states at $k_x=0$ or $k_y=0$ if we wish to avoid the scattering paradox of surface states at the junction between two TIs with opposite helicity. This topological phase transition has been characterized by a topological invariant produced by the mirror symmetry of the system in Ref.\ \onlinecite{taka}.

Also, we found that the tachyonlike dispersion presented in Ref.\ \onlinecite{apal} does not correspond to physical solutions. Rather, it is a consequence of implementing the wrong general solution at points on the $(E,k)$ grid where the secular equation (\ref{eq:lambda}) has four repeated roots. This degeneracy occurs independent of boundary conditions and the same complication arises for a vacuum interface where one can analytically show that these solutions are unphysical. From the comparison of Eq.\ \ref{eq:tachyon} with the numerical results in Fig.\ \ref{fig:tachyon}, and the fact that the tachyonic solutions vanish if we use the correct general solution at the degeneracy, we conclude that this system does not show any signature of tachyons.

\begin{acknowledgments}
The authors would like to thank Dr.\ O.\ Leenaerts for the helpful discussions. This work was supported by the Research Foundation Flanders (FWO). 
\end{acknowledgments}

\appendix* 

\section{General solution}

After the substitution $k_z\rightarrow -i\partial_z$, the Schr\"odinger equation $\left(\mathcal{H}-E\right)\phi=0$ is given by a coupled system of homogeneous second order differential equations that is solved with the ansatz $\bm{\phi}(z)\sim e^{\lambda z}\Psi(E,\bm{k})$, where $\Psi(E,\bm{k})$ is an eigenvector of the Hamiltonian. This yields a system of algebraic equations which have a nonzero solution if $|\mathcal{H}(\bm{k},-i\lambda)-E|=0$ which we can write as the square of a biquadratic equation in $\lambda$, given by
\begin{equation} \label{eq:lambda}
  \begin{aligned}
    &D_1D_2\lambda^4+\left(B^2+D_1(E-L_2)+D_2(E-L_1)\right)\lambda^2 \\
    &+(E-L_1)(E-L_2)-A^2k^2=0,
  \end{aligned}
\end{equation}
with
\begin{align*}
  L_{1,2}&=C_0\pm M_0+\left(C_2\pm M_2\right)k^2, \\
  D_{1,2}&=C_1\pm M_1.
\end{align*}
This gives four doubly degenerate $\lambda_\alpha(E,\bm{k})$ in general, and the eigenspace of $\lambda_\alpha$ is spanned by two eigenvectors $\Psi_{s,\alpha}(E,\bm{k})$ which can be chosen as \cite{shan}
\begin{equation} \label{eq:vec}
  \Psi_{1,\alpha}=\begin{pmatrix} -iB\lambda_\alpha \\ E-L_1+D_1\lambda_\alpha^2 \\ Ak_+ \\ 0 \end{pmatrix},
  \Psi_{2,\alpha}=\begin{pmatrix} 0 \\ Ak_- \\ E-L_2+D_2\lambda_\alpha^2 \\ iB\lambda_\alpha \end{pmatrix}.
\end{equation}
The general solution is then given by Eq.\ (\ref{eq:wf}). However, if the $\lambda$ are further degenerate, i.e. when Eq.\ (\ref{eq:lambda}) has a repeated root, there are only two distinct $\lambda$ that are both four times degenerate. In the last case, there are only two linear independent solutions for each $\lambda$, given by $e^{\lambda z}\Psi_1$ and $e^{\lambda z}\Psi_2$. In order to find two more linear independent solutions, we first consider a general system of $n$ homogeneous second order ordinary differential equations:
\begin{equation} \label{eq:ode}
  \bm{\phi}^{\prime\prime}=Q\bm{\phi}^\prime+P\bm{\phi},
\end{equation}
where $P$ and $Q$ are constant $n\times n$ matrices. If we plug $\bm{\phi}(z)=e^{\lambda z}\bm{\eta}$ in Eq.\ (\ref{eq:ode}), we find
\begin{equation} \label{eq:eta}
  \left(A+\lambda B-\lambda^2I\right)\bm{\eta}=0,
\end{equation}
which has a nonzero solution $\bm{\eta}\neq0$ if and only if $\det\left(A+\lambda B-\lambda^2I\right)=0$. In general, the solution of this equation is given by the roots of a polynomial of order $2n$ in $\lambda$. Now consider the case where there is a repeated root. If there are two linearly independent $\bm{\eta}$ corresponding to the repeated root, there is no problem. When this is not the case, we need to look for another linear independent solution. If we try $\bm{\phi}(z)=ze^{\lambda z}\bm{\eta}+e^{\lambda z}\bm{\rho}$, and equate terms of equal power in $z$, we find Eq.\ (\ref{eq:eta}) again and additionally we have
\begin{equation} \label{eq:rho} 
  \left(A+\lambda B-\lambda^2I\right)\bm{\rho}=\left(2\lambda I-B\right)\bm{\eta}.
\end{equation}
If we take the partial derivative of Eq.\ (\ref{eq:eta}) with respect to $\lambda$, and compare the result to Eq.\  (\ref{eq:rho}), we find $\bm{\rho}=\partial_\lambda\bm{\eta}$. If we use this result, we obtain the general solution when the $\lambda$ are four times degenerate:
\begin{equation} \label{eq:wf2}
  \bm{\phi}(z)=\displaystyle\sum\limits_{s,\beta=1,2}e^{\lambda_\beta z}\left[C_{s,\beta}\Psi_{s,\beta}+C_{s,\beta+2}\left(z+\partial_{\lambda_\beta}\right)\Psi_{s,\beta}\right].
\end{equation}


\begin{thebibliography}{99}

\bibitem{kane1} C. L. Kane and E. J. Mele, Phys. Rev. Lett. \textbf{95}, 146802 (2005).
\bibitem{bern} B. A. Bernevig, T. L. Hughes, and S. C. Zhang, Science \textbf{314}, 1757 (2006).
\bibitem{konig} M. K\"onig, S. Wiedmann, C. Br\"une, A. Roth, H. Buhman, L. W. Molenkamp, X. L. Qi, and S. C. Zhang, Science \textbf{318}, 766 (2007).
\bibitem{fu1} L. Fu, C. L. Kane, and E. J. Mele, Phys. Rev. Lett. \textbf{98}, 106803 (2007).
\bibitem{moore1} J. E. Moore and L. Balents, Phys. Rev. B \textbf{75}, 121306 (2007).
\bibitem{fu2} L. Fu and C. L. Kane, Phys. Rev. B \textbf{76}, 045302 (2007).
\bibitem{mura1} S. Murakami, N. J. Phys. \textbf{9}, 356 (2007).
\bibitem{hsieh} D. Hsieh, D. Qian, L. Wray, Y. Xia, Y. S. Hor, R. J. Cava, and M. Z. Hasan, Nature \textbf{452}, 970 (2008).
\bibitem{zhang} H. Zhang, C. X. Liu, X. L. Qi, X. Dai, Z. F. Fang, and S. C. Zhang, Nat. Phys. \textbf{5}, 438 (2009).
\bibitem{roy} R. Roy, Phys. Rev. B \textbf{79}, 195322 (2009).
\bibitem{xia} Y. Xia, D. Qian, D. Hsieh, L. Wray, A. Pal, H. Lin, A. Bansil, D. Grauer, Y. S. Hor, R. J. Cava, and M. Z. Hasan, Nat. Phys. \textbf{5}, 398 (2009).
\bibitem{chen} Y. L. Chen, J. G. Analytis, J. H. Chu, Z. K. Liu, S. K. Mo, X. L. Qi, H. J. Zhang, D. H. Lu, X. Dai, Z. Fang, S. C. Zhang, I. R. Fisher, Z. Hussain, and Z. X. Shen, Science \textbf{325}, 178 (2009).

\bibitem{kane2} C. L. Kane, Nat. Phys. \textbf{4}, 348 (2008).
\bibitem{qi1} X. L. Qi and S. C. Zhang, Phys. Today \textbf{63}, 33 (2010).
\bibitem{moore2} J. E. Moore, Nature \textbf{464}, 194 (2010).
\bibitem{hasan} M. Z. Hasan and C. L. Kane, Rev. Mod. Phys. \textbf{82}, 3054 (2010).
\bibitem{qi2} X. L. Qi and S. C. Zhang, Rev. Mod. Phys. \textbf{83}, 1057 (2011).
\bibitem{mura2} S. Murakami, N. J. Phys. \textbf{13} 105007 (2011).

\bibitem{taka} R. Takahashi and S. Murakami, Phys. Rev. Lett. \textbf{107}, 166805 (2011).
\bibitem{liu1} C. X. Liu, X. L. Qi, H. Zhang, X. Dai, Z. Fang, and S. C. Zhang, Phys. Rev. B \textbf{82}, 045122 (2010).


\bibitem{apal} V. M. Apalkov and T. Chakraborty, Eur. Phys. Lett. \textbf{100}, 17002 (2012).
\bibitem{shan} W. Y. Shan, H. W. Lu, and S.Q. Shen, N. J. Phys. \textbf{12}, 043048 (2012).
\bibitem{ben} D. J. BenDaniel and C. B. Duke, Phys. Rev. \textbf{152}, 683 (1966).
\bibitem{lob} I. Lobato and B. Partoens, Phys. Rev. B \textbf{83}, 165429 (2011).
\bibitem{dip} D. Sen and O. Deb, Phys. Rev. B \textbf{85}, 245402 (2012).
\bibitem{kats} M. I. Katsnelson, K. S. Novoselov, and A. K. Geim, Nat. Phys. \textbf{2}, 620 (2006).
\bibitem{brune} C. Br\"une, C. X. Liu, E. G. Novik, E. M. Hankiewicz, H. Buhmann, Y. L. Chen, X. L. Qi, Z. X. Shen, S. C. Zhang, and L. W. Molenkamp, Phys. Rev. Lett. \textbf{106}, 126803 (2011).
\bibitem{berch} N. N. Berchenko and M. V. Pashkovskii, Sov. Phys. Usp. \textbf{19}, 462 (1976).
\bibitem{liu2} C. X. Liu, T. L. Hughes, X. L. Qi, K. Wang, and S. C. Zhang, Phys. Rev. Lett. \textbf{100}, 236601 (2008).

\end{thebibliography}
\end{document}